# Changes in Heat Wave Characteristics over Extremadura (SW Spain)

Francisco Javier Acero, Departamento de Física, Universidad de Extremadura, Badajoz, Spain.

María Isabel Fernández Fernández, Departamento de Física, Universidad de Extremadura, Badajoz, Spain.

Víctor Manuel Sánchez Carrasco, Departamento de Física, Universidad de Extremadura, Badajoz, Spain.

Sylvie Parey, EDF/R&D, France.

Thi Thu Huong Hoang, EDF/R&D, France.

Didier Dacunha-Castelle, Laboratoire de Mathématiques, Université Paris 11, Orsay, France.

José Agustín García, Departamento de Física, Universidad de Extremadura, Badajoz, Spain.

**Corresponding author:** F.J. Acero, Departamento de Física, Universidad de Extremadura, 06006 Badajoz, Spain. (fjacero@unex.es)

**Abstract**

Heat wave (HW) events are becoming more frequent, and they have important consequences because of the negative effects they can have not only on the human population in health terms, but also on biodiversity and agriculture. This motivated a study of the trends in HW events over Extremadura, a region in the southwest of Spain, with much of its area in summer devoted to the production of irrigated crops such as maize and tomatoes. Heat waves were defined for the study as two consecutive days with temperatures above the 95th percentile of the summer (June-August) maximum temperature ($T_{max}$) time series. Two datasets were used: one consisted of 13 daily temperature records uniformly distributed over the Region, and the other was the SPAIN02 gridded observational dataset, extracting just the points corresponding to Extremadura. The trends studied were in the duration, intensity, and frequency of HW events, and in other parameters such as the mean, low (25th percentile), and high (75th percentile) values. In general terms, the results showed significant positive trends in those parameters over the east, the northwest, and a small area in the south of the Region. In order to study changes in HW characteristics (duration, frequency and intensity) considering different subperiods, a stochastic model was used to generate 1000 time series equivalent to the observed ones. The results showed that there were no significant changes in HW duration in the last 10-year subperiod in comparison with the first. But the results were different for warm events (WE), defined with a lower threshold (the 75th percentile), which are also important for agriculture. For several sites, there were significant changes in WE duration, frequency, and intensity.

# 1. Introduction

Heat wave (HW) events are exceptional extreme meteorological phenomena that are increasing not only in frequency, but also in duration and in intensity, as many studies have shown during the last two decades (Easterling et al. 2000; Klein Tank and Können 2003; Meehl and Tebaldi 2004; Schär et al. 2004; Klein Tank et al. 2005; Della-Marta 2007; Beniston et al. 2007; Acero et al. 2014; Russo et al. 2015). Also, events of the severity of the last two events in Europe in 2003 and 2010, classified as 'mega-heatwaves' by Barriopedro et al. (2011), are expected to occur with increasing frequency over highly populated areas of Europe. As a consequence of the Russian heat wave in 2010, tens of thousands of people died, with this event being the worst in Europe since at least 1950 (Russo et al. 2015). Also according to that study, regional projections suggest that in the next two decades (2021-2040) there will be heat wave events in Europe similar to or even more severe than that corresponding to the Russian in magnitude, extent, and duration (Russo et al. 2015).

According to the last Technical Report on Impacts, Adaptation, and Vulnerability developed by the Intergovernmental Panel on Climate Change (IPCC 2014): "impacts from recent climate-related extremes, such as heat waves, droughts, floods, cyclones, and wildfires, reveal the significant vulnerability and exposure of some ecosystems and many human systems to current climate variability".

The study area of the present work is the Region of Extremadura in southwestern Spain. It is characterized by its dependence on agriculture, particularly the production of irrigated crops in summer, and by its biodiversity which is an important aspect attracting tourism. The agricultural sector in Extremadura contributes 53% to the Region's total Gross Domestic Product (GDP) directly or indirectly. This high value is mainly due to the existence of a major proportion of agro-food industries. Many studies have demonstrated the influence of climate change on agriculture. For example, the aforementioned IPCC Report analyses the impacts of the 2003 HW on agriculture (IPCC, 2014). Maize production in the Po Valley in Italy fell by 36%, and in France, compared to the previous year (2002), there was a 30% reduction in the maize harvest and 20% in fruit production (Ciais et al. 2005). Schlenker et al. (2009) studied the damage to crop yields of the nonlinear temperature effects of climate change. The results showed that, due to the increase in maximum temperatures, area-weighted average yields are predicted to decrease by 30–46% before the end of the century under the slowest warming scenario (B1), and by 63– 82% under the fastest warming scenario (A1FI) according to a climate model (Schlenker et al. 2009).

As mentioned above, there have been many studies in the last two decades showing a general tendency towards higher maximum temperature extremes in Europe. Different studies have considered the Iberian Peninsula (IP) as a whole, just Spain, or an area inside Spain using a variety of indices taken from the set provided by the Expert Team on Climate Change Detection and Indices (ETCCDI) recommended by the World Meteorological Organization (Zhang et al. 2011) to investigate trends in maximum temperatures. A spatial regionalization of summer temperature extremes has been developed for the northeast of the IP (Kenawy et al. 2013). Also, according to a study of non-urban observatory data over the IP (Acero et al. 2014), temperature extremes are expected to increase but not as much as the mean temperatures because the variance tends to decrease.

The aim of the present work is to study trends in HW characteristics: duration, frequency, and intensity. For this purpose, a set of 13 regularly distributed daily maximum temperature ($T_{max}$) time series was selected from a larger database covering the Region of Extremadura in the southwestern IP. The SPAIN02 gridded observational dataset (Herrera et al. 2012, 2015) for $T_{max}$ was also used, considering three horizontal resolutions (0.11°, 0.22°, and 0.44°), and selecting just those grid points corresponding to the study region. Apart from HW events defined as exceedances over the 95th percentile, we also studied Warm Events (WE) defined as exceedances over the 75th percentile because of their important consequences on agriculture (Hatfield et al 2011). This study of the evolution of HW characteristics was based on identifying trends in both the observed temperature time series and simulated temperature time series consistent with the observed ones.

The paper is organized as follows. The data that were selected and analysed are described in Sec. 2. The methodology used is analysed in Sec. 3. The results are presented and discussed in Sec. 4. Finally, the conclusions are drawn in Sec. 5.

## 2. Data

The study area was the Extremadura Region, in the southwest of the IP (Fig. 1), with a total area of 41 635 km². There is a contrasting orography between the wide areas of the Rivers Tagus and Guadiana depressions with altitudes under 400 m a.s.l and the highest peak of over 2400 m a.s.l. The northern area borders the Northern Meseta, and has greater altitudes than the western, eastern, and southern limits.

The time series were drawn from an extensive daily temperature time series database provided by Spain's State Meteorological Agency (Agencia Estatal de Meteorología, AEMET) for use within the framework of the present study. A group of series had to be chosen that covered Extremadura's orographic diversity. The final choice was a set of 13 homogeneous daily temperature time series corresponding to sites as uniformly spaced over the Region as possible. Table 1 lists the period covered for each site, and their locations are shown in Fig. 1. There are no sites in the mountainous areas of Extremadura due to the absence of population and difficult accessibility. The altitudes of the sites chosen range from 185 to 796 m a.s.l. We also used the gridded observational temperature dataset SPAIN02 (Herrera et al. 2012, 2016) at three horizontal resolutions: 0.11°, 0.22°, and 0.44°. SPAIN02 (Herrera et al. 2012, 2015) is a series of high-resolution daily precipitation and (maximum and minimum) temperature gridded datasets developed for peninsular Spain and the Balearic Islands. A dense network of ~2500 quality-controlled stations (~250 for temperatures) for the period 1950-2007 was selected from the data of Spain's State Meteorological Agency (AEMET) in order to build the gridded products for version v2 (0.2° regular grid). The updated version v4 includes three resolutions (0.11°, 0.22°, and 0.44° in rotated coordinates) for the period 1971-2007. In the work of Herrera et al. (2016), both mean and extreme indicators were considered to show that both regimes are correctly reproduced by the gridded dataset. This gridded dataset was used to check the main trends in $T_{max}$ with the observational data in the common period 1971-2007.

The homogeneity of the monthly time series data was evaluated using the R-based software package RHtestsV3, developed and maintained at the Climate Research Division of Canada's Meteorological Service, and available from the ETCCDI website (http://etccdi.pacificclimate.org/software.shtml). This software is based on a two-phase regression model with a common linear trend (Wang 2003), and can identify multiple step changes at documented or undocumented change-points. This same procedure has been used in previous work (Acero et al. 2011, 2012, 2014). The results of the analysis, together with the sites' metadata, showed that all of the 13 time series were homogeneous in their respective periods of study, with none of them having change-points significant at 5%.

## 3. Methodology

### 3.1 Heat wave definition

There are lots of definition about heat wave involving temperature and other magnitudes such as humidity (Robinson 2001) Perkins, S.E. and L.V. Alexander, 2013: On the Measurement of Heat Waves. J. Climate, 26, 4500-4517, doi: 10.1175/JCLI-D-12-00383.1. Perkins and Alexander 2013). Considering, for example, the definition given by the National Oceanic and Atmospheric Administration (NOAA): "a period of abnormally and uncomfortably hot and unusually humid weather. Typically a heat wave lasts two or more days". But the study region is characterized by very dry summers leading to a non significative use of the humidity in order to define the heatwave. Then, the heat wave is defined as two consecutive days with temperature over a high threshold. Besides, since the object of the study is the occurrence of heat waves, the threshold that we applied, $u$, is the definition of heat wave given by Spain's General Directorate of Civil Protection and Emergency belonging to the Department of the Interior, which is responsible for activating weather alerts for the population. This definition is the 95th percentile of the summer temperature time series. A more demanding definition, three consecutive days over the aforementioned threshold $u$, was also considered but the results lead to a very low number of heatwaves over the whole period disregarding this definition.

Then, heat waves are defined as two consecutive days occurring above a threshold of the summer (June-August) maximum temperature time series. The threshold ($u$) used was the 95th percentile of the daily $T_{max}$ for both the observational and the three gridded datasets. To illustrate the data used, Fig. 2 shows the spatial distribution of this threshold $u$ for these datasets in Extremadura. The patterns of the threshold for the three gridded datasets are quite similar, with higher values in the southeast corresponding to the hottest areas reaching nearly 40°C, and lower in the north, corresponding to the most mountainous area with values of around 38°C. That the values of the gridded data scale are lower is due to the influence of the northern sites used in the kriging procedure. These correspond to a part of Extremadura in Spain's Northern Plateau where the values of $T_{max}$ are lower due to the higher altitude.

Fig. 3 shows that the trends in the mean maximum temperature are positive and significant for the majority of the sites. Hence, there might also be adverse effects for agriculture (Hatfield et al 2011) due to an increase in events not as severe as a HW, but still being episodes of consecutive days exceeding a lower, although high, threshold. The pollinitation stage is one of the more susceptible phenological stages and, for example, maize pollen viability decreases with exposure to temperatures above 35 °C (Herrero and Johnson, 1980). For that reason, warm events had been defined as exceedances over the 75th percentile which, for the study region, range from 34°C in the north to 37°C in the southeast.

**3.2 Trend identification and testing**

3.2.1 Statistical trend tests

The HW duration, intensity, and frequency were estimated for each year, and the resulting values were used to estimate the significance of the trend by means of a two-sided Mann-Kendall (M-K) test (Kendall 1976) and the value of the trend with the Theil-Sen estimator (Sen 1968). For each year, the duration was taken to be the mean HW duration in days, the frequency to be the number of events in that year, and the intensity the mean value of the sum of the exceedances over the threshold for each HW.

The Mann-Kendall test allowed monotonic (increasing or decreasing) trends in these variables (HW duration, frequency, and intensity) to be identified. One of the main advantages of this test is that it is non-parametric and therefore does not require the data to be normally distributed. In order to estimate the magnitude of the trends, we used the algorithm proposed by Hirsch and Smith (1982) which extends the original test put forward by Theil (1950a, 1950b, 1950c) and Sen (1968). The statistics of this Theil-Sen test are related to the slope of the trend found by the M-K test. This procedure has been extensively used for trend estimations (e.g., Gallego et al., 2006, 2011).

However there is a limitation in the results given by the M-K test because, firstly, a real trend may not be monotonic (as was indeed the case, since sometimes the second decade was cooler than the first), and secondly, it does not take into account the variability around the means that it tests and observation gives just one realization of the variables. In order to go more deeply into the analysis, a simulation technique was used to produce a large number of equivalent time series and thus allow for more robust statistics.

3.2.2 Stochastic modelling

The simulation was based on a recently developed stochastic seasonal functional heteroskedastic autoregressive model capable of simulating daily minimum, maximum, or mean temperature time series that are coherent with the observed series (Parey et al. 2014; Dacunha-Castelle et al., 2015). It was designed to reliably reproduce extreme values. The model simulates $Z(t) = \frac{X(t) - s_m(t) - m(t)}{s_v(t) - s(t)}$, where $X(t)$ is the observed temperature time series, $s_m(t)$ and $s_v(t)$ are the seasonalities in the mean and the standard deviation, and $m(t)$ and $s(t)$ the trends in mean and standard deviation, respectively. The seasonalities are estimated as trigonometric polynomials of the form, $\theta_0 + \sum_{i=1}^{p} \left( \theta_{i,1} \cos \frac{2i\pi t}{365} + \theta_{i,2} \sin \frac{2i\pi t}{365} \right)$, where the order $p$ is chosen according to an Akaike criterion (Parey and Hoang, 2016). For the trends in mean and variance, non-parametric estimation by LOESS fitting was chosen. The smoothing parameter needed was chosen by means of a modified partitioned cross-validation technique (Parey and Hoang, 2016). More details of the model are given in Dacunha-Castelle et al. (2015). Validation of the model in different climatic regions in Eurasia and the United States (Parey et al. 2014) showed it to produce coherent results for both the bulk of a distribution and its extremes.

In the present work, this stochastic temperature generator was used to produce 1000 time series equivalent to the observed ones in order to investigate the significance of the changes in HW characteristics by applying subperiods of different lengths covering the observed period. Different 10-year subperiods were considered within the observed period. The mean durations, frequencies, and intensities were calculated for the simulated data for each subperiod, and compared with the first subperiod as reference (1960-1969 for the time series beginning in 1960, and 1965-1974 for those beginning in 1965). A characteristic was taken to have changed if its value in the last period lay outside the 90% CI of its value in the reference period.

The ability of the model to simulate realistic heat wave characteristics for the considered region has been checked (see supplementary material).

## 4. Results

### 4.1 Heat Waves

Figure 4 shows the spatial distribution of the trends in HW duration, intensity, and frequency. The pattern for all three is quite similar, with positive trends significant at a 5% significance level for the west, the east, and a small area in the south. Also, there is good agreement between the results for the observed and the gridded datasets. One observes that the frequency has a smaller area with significant trends than the duration and intensity, with there being an extensive area without any significant trend in the number of HW events.

Then the stochastic model was used to analyse the changes in the HW characteristics of duration, frequency, and intensity. Table 2 summarizes which sites presented significant changes in WE or HW characteristics. With respect to HW duration, no significant differences were found in comparing the last subperiod with the reference subperiod. The greatest differences corresponded to three sites (cor, cij, svi), but they did not reach significance at the 10% level. Since the first subperiod was warmer than the second for the time series beginning in 1960, this second subperiod was also tested as reference instead of the first. But again, no significant differences were found. No changes were significant even when subperiods of 5 years were tested, although the HW duration had increased by the last subperiod. By way of illustration, Fig. 5 (top four figures) shows the results obtained for San Vicente de Alcántara. This site shows an increase in HW duration, but, even considering the last period, the values remain within the reference period's CI. For the rest of observatories, the duration of the HW events is increasing from the first subperiod (1960-1969) to the last (2000-2014), but not significantly.

The results were different for the HW frequencies, with 7 sites (bad, ber, cac, cij, cor, gua, svi) showing significant changes at the 10% significance level in comparing the last subperiod with the reference. Figure 5 (four bottom figures) shows the changes in HW frequency for Coria. The Coria site showed significant changes in 1985-1994, 1995-2004, and 2005-2014 compared to the reference subperiod of 1965-1974.

As was the case with duration, there were no significant changes in HW intensity for any of the sites studied.

Thus although significant monotonic trends had been found according to the M-K test, increasing the variability by use of the stochastic generator showed that only the frequency of HW may have significantly changed.

**4.2 Warm events**

Figure 6 shows the spatial distribution of the trends in WE duration, intensity, and frequency. In general, all three of the variables present positive trends for the whole of Extremadura. Comparing this figure with the figure 4 for HW events, one observes that intensity is the variable with the greatest increase for WE in both the values of the trends and the area with a significant positive trend at the 5% significance level. As illustration of this behaviour, the site of San Vicente de Alcántara (svi), the westernmost one, shows a positive trend significant at the 5% significance level of magnitude 0.55°C per decade for HW events but of magnitude 2.61°C per decade for WE events. There is also an increase in the trend in frequency for this site from 0.33 HW events per decade to 1.02 WE per decade in changing the threshold from the 95th to 75th percentile. The comparison of Figs. 4 and 6 shows that, although there are positive trends in many areas of Extremadura in frequency, intensity, and duration of the HW events, for WE the trends are stronger in these three variables, implying negative consequences for a Region such as this based on farming and characterized by its biodiversity.

As noted above, climate variability plays an important role, with the 1970s being cooler than the 1960s. Therefore, for each site, the coolest 10-year subperiod was chosen as reference against which to compare the duration of these WE when using the stochastic modelling. The results showed 4 sites (cij, cor, jer, svi) to have significant changes in these WE durations of the last subperiod compared to the reference. These changes are illustrated in Fig. 7 for San Vicente de Alcántara (svi).

The analysis of changes in WE frequency showed 8 sites to have significant changes in the last subperiod compared with the reference one. 3 sites (cij, cor, svi) showed significant changes in WE intensity.

Thus more significant changes were found for WE than for HW. This may be linked to the large increase in the lower $T_{max}$ values over the region, as shown in Fig. 8 for the changes in the 25th percentile of the summer $T_{max}$ distribution compared to those corresponding to the 75th percentile.

## 5. Conclusions

Since many authors have described important negative impacts of HW events on agriculture and biodiversity, we undertook a study of HW events in Extremadura (SW Spain), calculating the trends in changes in HW characteristics (duration, frequency, and intensity) mainly using statistical tools and a stochastic model for temperatures. We also defined Warm Events (WE) so as to study trends in less severe warm temperatures. Two datasets were used: one was a set of complete daily $T_{max}$ time series from 13 sites for the period 1960-2014, and the other was the gridded dataset SPAIN02 for the period 1971-

2007.

The HW events were defined as two or more consecutive days above the 95th percentile of summer maximum temperatures. In general, trends were found in HW duration, frequency, and intensity over the study area that were significant at a 5% significance level according to the Mann-Kendall test. However, this test only looks for monotonic trends, which is a strong assumption since interannual variability leads to the behaviour being nonlinear. Besides, observation gives just one realization of the variables, and it does not allow internal climate variability to be taken into account. Therefore, a stochastic model was used to generate 1000 maximum temperature time series equivalent to the observed series in order to go more deeply into the analysis of changes in the duration, frequency, and intensity of HW events in the observed period. For this purpose, we considered 10-year subperiods, taking the coolest as reference against which to compare the characteristics of the other subperiods. This analysis found no significant changes in HW duration and intensity, since the values corresponding to the last observed 10-year subperiod lay within the 90% CI's of the reference period. But there were significant increases in the frequency of HW events for 7 sites in the last 15 years of the study period.

As well as HW events, we studied warm events (WE) defined as two or more consecutive days above the 75th percentile of the summer maximum temperature time series. In Extremadura, this value ranges from 34.5°C in the north to 37°C in the southeast. It is of especial interest in an agricultural area such as this for which the summer season is crucial for crops such as tomatoes, maize, and rice. Although this is not a threshold that defines very severe events, our aim was to analyse whether the trends in the duration, frequency, and intensity of these warm events were increasing more than those of HW, and to study any significant changes found in the WE characteristics of different subperiods using the stochastic model applied to the observed time series. The results showed the trends in WE frequency and intensity to be stronger than those corresponding to HW. There were significant changes in the last 10-year subperiod compared to the first subperiod for 4 of the 13 sites in WE duration, 3 of the sites in intensity, and 8 in frequency. Thus, while the parameters of the HW events presented a moderate increase, those corresponding to WE increased more strongly. This is linked to hot summers not only with longer and more intense events of temperatures above 35-36°C, but also fewer days with milder temperatures as reflected in the marked increase in low values (25th percentile) of $T_{max}$. Therefore, different classes of events such as those analysed in the present work should be considered in order to study the effects on agriculture and biodiversity under the conditions of climate change, like the aforementioned decrease of the maize pollen viability at temperatures higher than 35ºC, widely exceeded in the study region.

**Acknowledgements**

Thanks are due to the Spanish State Meteorological Agency (Agencia Estatal de Meteorología: www.aemet.es) for providing the daily temperature time series used in this study. This work was partially supported by the Junta de Extremadura-FEDER Funds through Research Project (IB13049) and Research Group Grants (GR15137).

**References**

Acero FJ, García JA, Gallego MC (2011) Peaks-over-threshold study of trends in extreme rainfall over the Iberian Peninsula. J Clim 24:1089-1105. doi: 10.1175/2010JCLI3627.1

Acero FJ, Gallego MC, García JA (2012) Multi-day rainfall trends over the Iberian Peninsula. Theor Appl Climatol 108:411-423. doi 10.1007/s00704-011-0534-5 24

Acero FJ, García JA, Gallego MC, Parey S, Dacunha-Castelle D (2014) Trends in summer extreme temperatures over the Iberian Peninsula using nonurban station data. J Geophys Res 119:39–53. doi:10.1002/2013JD020590

Acero FJ, Parey S, Fernández-Fernández MI, Carrasco VMS, García JA (2016) http://meetingorganizer.copernicus.org/EGU2016/EGU2016-6036.pdf.

Barriopedro D, Fischer EM, Luterbacher J, Trigo RM, García-Herrera R (2011) The hot summer of 2010: redrawing the temperature record map of Europe. Science 332:220-224.

Beniston M, Stephenson DB, Christensen OB, Ferro CAT, Frei C, Goyette S, Halsnaes K, Hollt T, Jylhä K, Koffi B, Palutikof J, Schöll R, Semmler T, Woth K (2007) Future extreme events in European climate: an exploration of regional climate model projections. Clim Change 81:71–95.

Ciais, P. et al. (2005) Europe-wide reduction in primary productivity caused by the heat and drought in 2003. Nature 437:529-533

Dacunha-Castelle D, Hoang TTH, Parey S (2015) Modeling of air temperatures: preprocessing and trends, reduced stationary process, extremes, simulation. Journal de la Société Française de Statistique 156:138-168.

Della-Marta PM, Haylock MR, Luterbacher J, Wanner H (2007) Doubled length of western European summer heat waves since 1880. J Geophys Res 112, D15103. doi10.1029/2007JD008510.

Easterling DR, Meehl GA, Parmesan C, Changnon SA, Karl TR, Mearns LO (2000)

Climate extremes: observations, modelling, and impacts. Science 289:2068-2074. doi: 10.1126/science.289.5487.2068

Gallego MC, García JA, Vaquero JM, Mateos VL (2006) Changes in frequency and intensity of daily precipitation over the Iberian Peninsula. J Geophys Res 111, D24105, doi:10.1029/ 2006JD007280

Gallego MC, Trigo RM, Vaquero JM, Brunet M, García JA, Sigró J, Valente MA (2011) Trends in frequency indices of daily precipitation over the Iberian Peninsula during the last century. J Geophys Res 116, D02109. doi:10.1029/2010JD014255

Hatfield JL, Boote KJ, Kimball BA, Ziska LH, Izaurralde RC, Ort D, Thomson AM, Wolfe DW (2011) Climate impacts on agriculture: implications for crop production. Agron J 103:351-370.

Herrera S, Gutiérrez JM, Ancell R, Pons MR, Frías MD, Fernández J (2012) Development and analysis of a 50-year high-resolution daily gridded precipitation dataset over Spain (Spain02). Int J Climatol 32:74-85. doi: 10.1002/joc.2256

Herrera S, Fernández J, Gutiérrez JM (2016) Update of the Spain02 gridded observational dataset for EURO-CORDEX evaluation: assessing the effect of the interpolation methodology. Int J Climatol 36:900–908. doi:10.1002/joc.4391

Herrero MP, Johnson RR (1980) High temperature stress and pollen viability in maize. Crop Sci 20: 796-800.

IPCC, 2014: Summary for policymakers. In: Climate Change 2014: Impacts, Adaptation, and Vulnerability. Part A: Global and Sectoral Aspects. Contribution of Working Group II to the Fifth Assessment Report of the Intergovernmental Panel on Climate Change [Field, C.B., V.R. Barros, D.J. Dokken, K.J. Mach, M.D. Mastrandrea,

T.E. Bilir, M. Chatterjee, K.L. Ebi, Y.O. Estrada, R.C. Genova, B. Girma, E.S. Kissel, A.N. Levy, S. MacCracken, P.R. Mastrandrea, and L.L. White (eds.)]. Cambridge University Press, Cambridge, United Kingdom and New York, NY, USA, pp. 1-32.

Kenawy AE, López-Moreno JI, Brunsell NA, Vicente-Serrano SM (2013) Anomalously severe cold nights and warm days in northeastern Spain: Their spatial variability, driving forces and future projections. Global Planet Change 101: 12–32.

Kendall S (1976) Time Series. Oxford Univ. Press, New York.

Klein Tank AMG, Können GP (2003) Trends in indices of daily temperature and precipitation extremes in Europe, 1946–99. J Clim 16:3665–3680.

Klein Tank AMG, Können GP, Selten FM (2005) Signals of anthropogenic influence on European warming as seen in the trend patterns of daily temperature variance. Int J Climatol 25:1–16.

Meehl GA, Tebaldi C (2004) More intense, more frequent, and longer lasting heat waves in the 21st century. Science 305:994-997.

Parey S, Hoang TTH, Dacunha-Castelle D (2014) Validation of a stochastic temperature generator focusing on extremes, and an example of use for climate change. Clim Res 59:61-75. doi:10.3354/cr01201

Parey S, Hoang TTH (2016) Changes in the distribution of cold waves in France in the middle and end of the 21st century with IPSL-CM5 and CNRM-CM5 models. Clim Dyn 47:879-893. doi:10.1007/s00382-015-2877-6

Perkins SE, Alexander LV (2013) On the measurement of heat waves. J Climate 26: 4500-4517. doi: 10.1175/JCLI-D-12-00383.1.

Robinson, PJ (2001) On the definition of a heat wave. J Appl Meteor 40: 762-775.

Russo S, Sillmann J, Fischer EM (2015) Top ten European heatwaves since 1950 and their occurrence in the coming decades. Environ Res Lett 10:124003. doi:10.1088/1748-9326/10/12/124003

Schär C, Vidale PL, Lüthi D, Frei C, Häberli C, Liniger MA, Appenzeller C (2004) The role of increasing temperature variability in European summer heatwaves. Nature 432:610-614.

Schlenker W, Roberts MJ (2009) Nonlinear temperature effects indicate severe damage to U.S. crop yields under climate change. Proceedings of the National Academy of Science of the United States 106:15594-15598.

Sen PK (1968) Estimates of regression coefficient based on Kendall's tau. J Am Stat Assoc 63:1369–1379.

Wang XL (2003) Detection of undocumented changepoints: A revision on the two-phase regression model. J Clim 16:3383–3385.

Zhang XF, Alexander L, Hegerl GC, Jones P, Klein Tank AMG, Peterson TC, Trewin B, Zwiers FW (2011) Indices for monitoring changes in extremes based on daily temperature and precipitation data. WIREs Clim Change 2: 851–870, doi:10.1002/wcc.147.

Supplementary material: validation of the stochastic model

To validate these simulations, different tests were performed: comparison of the moments, Q-Q plots, and various quantiles. Figure 9 shows the Q-Q plots for the observed and simulated values of $T_{max}$ at San Vicente de Alcántara (svi). The results were similar for the rest of the sites, confirming that the model reasonably reproduces the temperature distributions.

With regard to the entire distribution, Fig. 10 shows the distributions of different percentiles (1%, 10%, 50%, 60%, 90%, and 99%) obtained from the 1000 simulations (in black) and the values from the observations (in red) for San Vicente de Alcántara. The results show that, for all the percentiles, the observed estimates fall within the 95% confidence interval (CI) of the simulated estimates. The results were similar for the rest of the sites.

In order to check the ability of the model to produce HW events, we defined such events as periods of consecutive days with $T_{max}$ above the 99th percentile of the annual time series, which corresponds to the 95th percentile of the summer time series that had been chosen as threshold. For each site, the 99th percentile was computed, and the distribution of episodes of from 1 to more than 15 days in the observed time series was compared to the minimum, maximum, and mean frequencies of that distribution in the 1000 simulations. Figure 11 shows the results for HW events at San Vicente de Alcántara. The stochastic model suitably reproduces short and long events for this site, but it tends to overestimate the proportion of 1-day events. The results were similar for the remaining sites. Therefore, HW and WE will be defined as events of at least 2 days with temperatures above the chosen threshold, as was done in the previous section.

| Code | Name | Geographic coordinates (deg) | Length |
|------|------|------------------------------|--------|
| alc | Alcuéscar | 39.1806°N 6.2286°W | 1960-2014 |
| alm | Almendralejo | 38.69°N 6.3611°W | 1960-2014 |
| bad | Badajoz/Talavera | 38.8833°N 6.8139°W | 1960-2014 |
| bdo | Barrado | 40.0833°N 5.8825°W | 1960-2014 |
| ber | Berlanga | 38.2833°N 5.8325°W | 1960-2014 |
| cac | Cáceres | 39.4714°N 6.3389°W | 1960-2014 |
| cij | Cíjara Dam | 39.0372°N 5.0228°W | 1965-2014 |
| cor | Coria | 39.98°N 6.59°W | 1965-2014 |
| gab | Gabriel y Galán Dam | 40.2208°N 6.1256°W | 1965-2014 |
| gua | Guadalupe | 39.455°N 5.3333°W | 1960-2014 |
| jer | Jerez de los Caballeros | 38.3186°N 6.7714°W | 1960-2014 |
| svi | San Vicente de Alcántara | 39.3628°N 7.1367°W | 1965-2014 |
| vil | Villanueva de la Serena | 38.98°N 5.80°W | 1960-2014 |

Table 1. Code used in the map, name, geographic coordinates, and length of the period studied.

|      | WARM EVENTS (u=75th perc.) | | | HEAT WAVES (u=95th perc.) | | |
| :---: | :---: | :---: | :---: | :---: | :---: | :---: |
| Code | Duration | Frequency | Intensity | Duration | Frequency | Intensity |
| alc |   |   |   |   |   |   |
| alm |   |   |   |   |   |   |
| bad |   | X |   |   | X |   |
| dbo |   |   |   |   |   |   |
| ber |   | X |   |   | X |   |
| cac |   | X |   |   | X |   |
| cij | X | X | X |   | X |   |
| cor | X | X | X |   | X |   |
| gab |   |   |   |   |   |   |
| gua |   | X |   |   | X |   |
| jer | X | X |   |   |   |   |
| svi | X | X | X |   | X |   |
| vil |   |   |   |   |   |   |

Table 2. Sites (marked with X) with significant changes in WE or HW characteristics.

**Fig. 1** Location of the study area (Extremadura) in the Iberian Peninsula (top), the spatial distribution of the sites used (bottom-left), and the gridded dataset SPAIN02 for the resolution 0.11° (bottom-right).

**Fig. 2** Spatial distribution of the 95th percentile of maximum temperature (°C) used as threshold for the observational data (top-left) and for the three resolutions of the gridded SPAIN02 dataset.

**Fig. 3** Spatial distribution of the trends in the mean maximum temperature (°C per decade), as estimated by the Sen method. Black filled, grey filled, and open triangles mean trends that are significant at a 5% level of significance, significant only at a 10% level, and not significant, respectively, as determined by the two-sided Mann-Kendall test. Upward pointing triangles mean a positive trend, and downward pointing, negative. Open circles mean sites with a trend whose value is less than 15% of the highest trend.

**Fig. 4** Spatial distribution of the trends in duration in days per decade (top), frequency in number of heat waves per decade (middle), and intensity in °C per decade (bottom), for the heat wave events over Extremadura. Black filled, grey filled, and open triangles mean trends that are significant at a 5% level of significance, significant only at a 10% level, and not significant, respectively, as determined by a two-sided Mann-Kendall test. Upward pointing triangles mean a positive trend, and downward pointing, negative. The size of the triangles represents the value of the trend for each site. Open circles mean sites with a trend whose value is less than 15% of the highest trend.

**Fig. 5** Heat wave durations considering different subperiods for San Vicente de Alcántara (top four figures) and heat wave frequencies for Coria (bottom four figures) sites. The threshold used is the 95th percentile of the summer time series for the whole period. Solid black (blue) vertical lines correspond to the mean duration or the mean frequency for the reference (studied) period, and dashed lines correspond to the 90% CI.

**Fig. 6** As Fig. 4, but showing the spatial distribution of the warm events considered as exceedances over the 75th percentile defined as threshold.

**Fig. 7** Warm events durations for different subperiods for the San Vicente de Alcántara site. Solid black (blue) vertical lines correspond to the mean duration for the reference (studied) period, and dashed lines correspond to the 90% CI.

**Fig. 8** Spatial distribution of the trends in the low (25th percentile – top) and high (75th percentile – bottom) values of the summer maximum temperature in °C per decade for the observational (left) and gridded (right) datasets.

**Fig. 9** Q-Q plots of the daily maximum temperature at San Vicente de Alcántara (svi). The solid line means a diagonal (i.e., 1:1) relationship; open circles are for the simulations, and the dashed lines show the 95% confidence interval of the simulations.

**Fig. 10** Distributions (solid blue lines) of the 1st, 10th, 50th, 60th, 90th, and 99th percentiles of the 1000 simulated maximum temperature distributions; dashed lines show the 95% confidence interval (2.5% and 97.5% quantiles of the 1000 simulations) estimated for the maximum temperature at the San Vicente de Alcántara (svi) site. The red lines correspond to the same percentile determined from the observations.

**Fig. 11** Frequencies of the 2 to >15 days length heat waves at San Vicente de Alcántara (svi). Red lines are the observed frequencies. Solid black lines are the mean frequencies obtained from the simulations, and the dashed lines are the corresponding minimum and maximum frequencies;

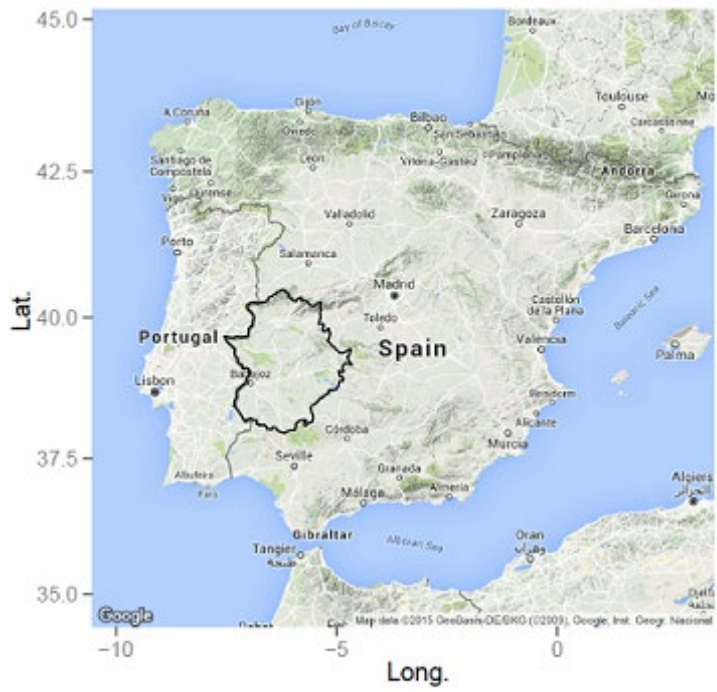
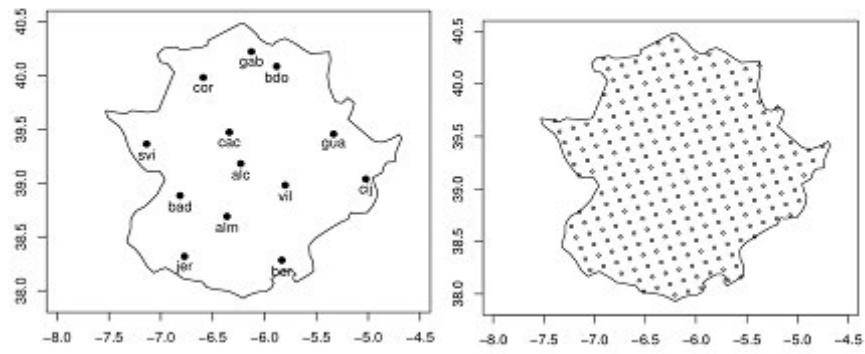

Observational data 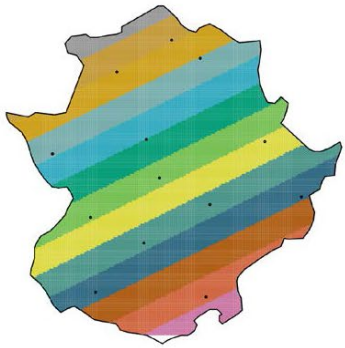 Gridded data - 0.44° 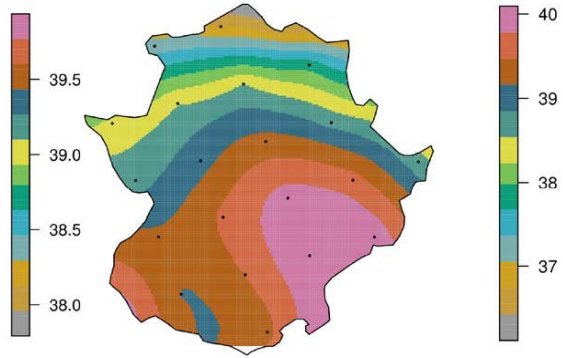

Gridded data - 0.22° 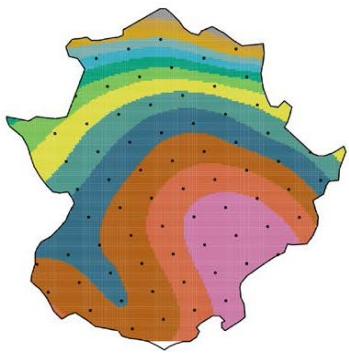 Gridded data - 0.11° 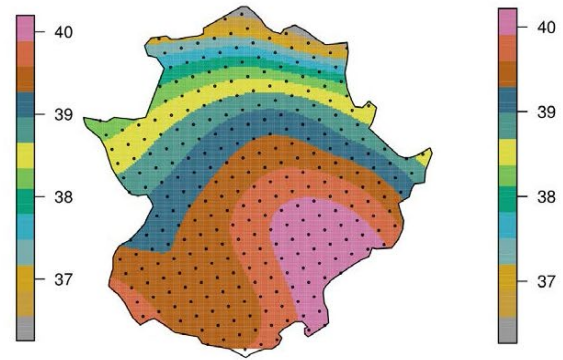

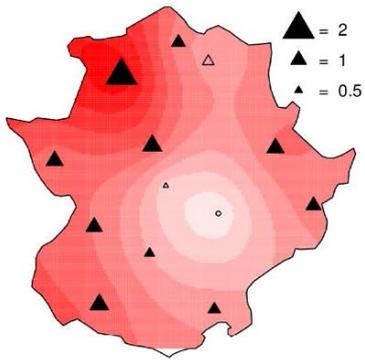 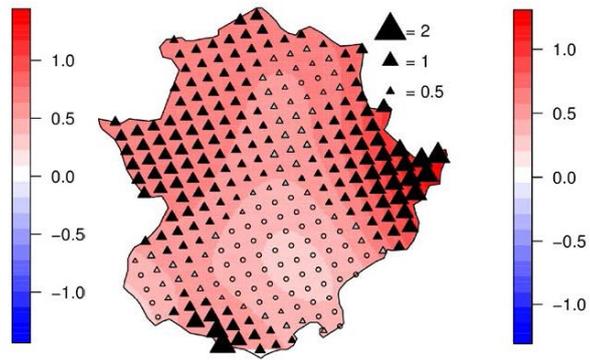

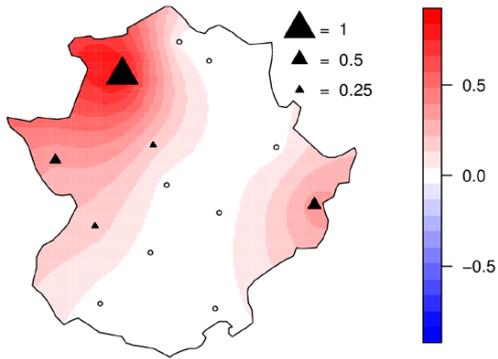
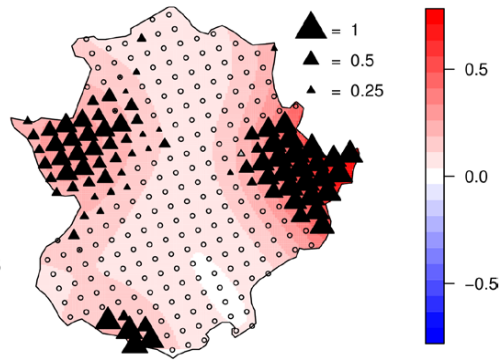
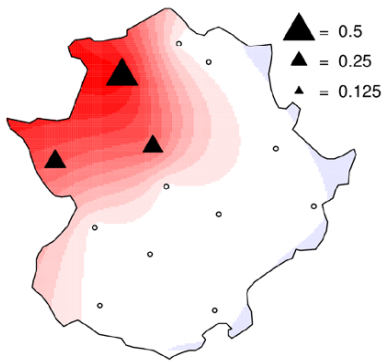
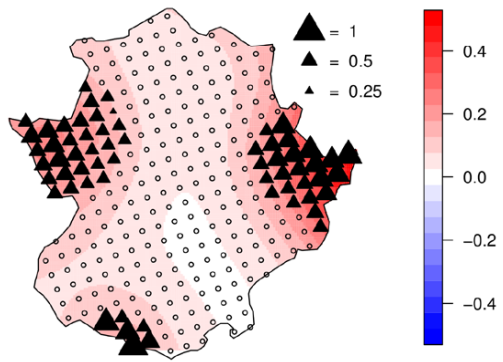
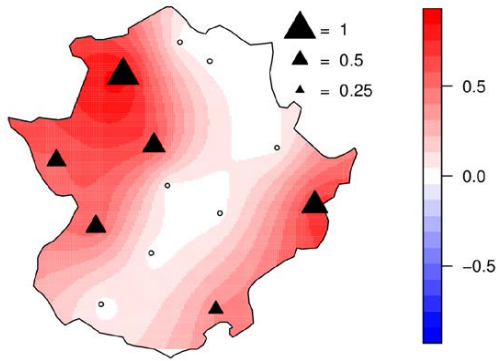
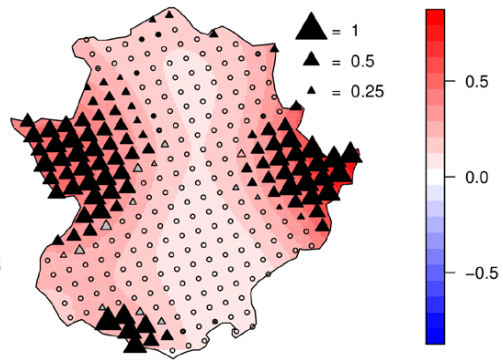

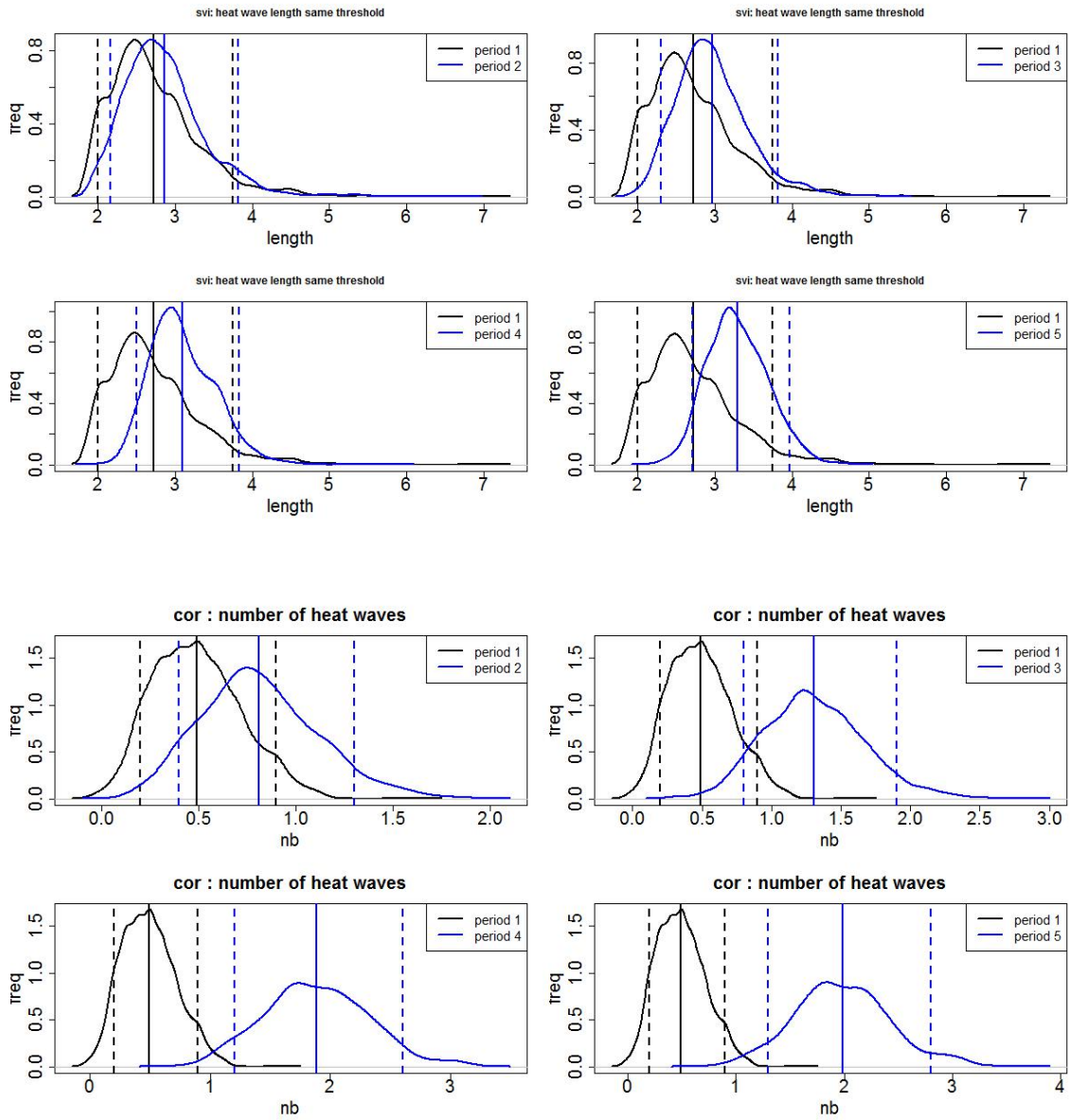

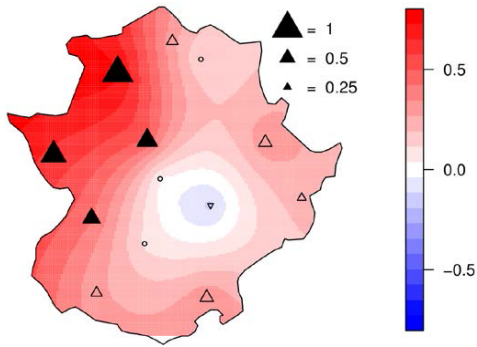
Trends in WE duration (observational data)

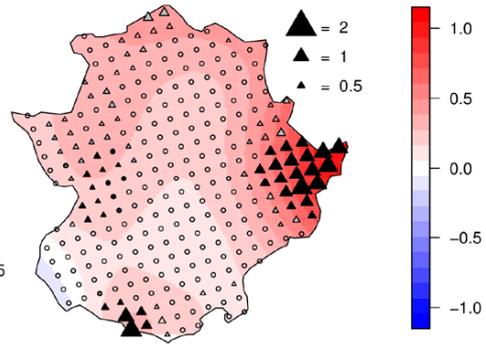
Trends in WE duration (gridded data)

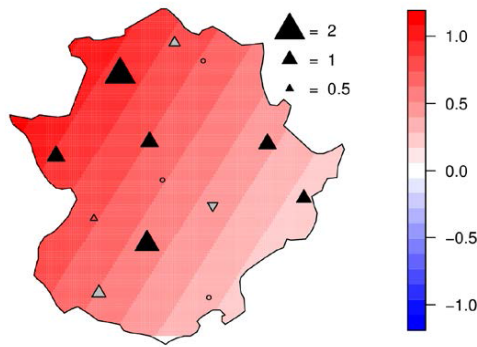
Trends in WE frequency (observational data)

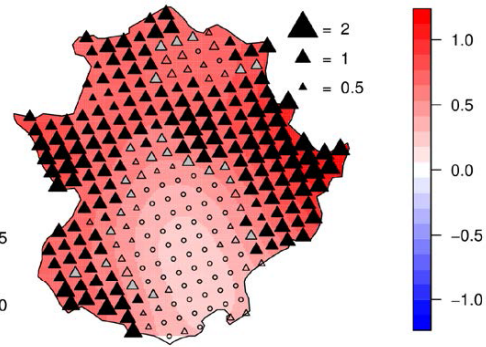
Trends in WE frequency (gridded data)

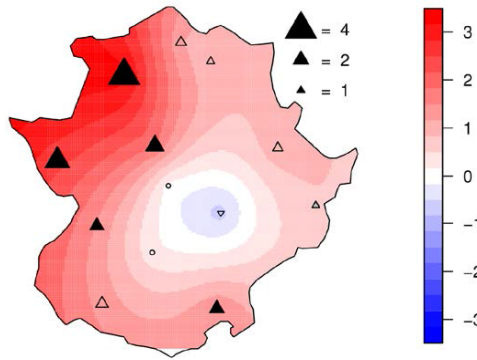
Trends in WE intensity (observational data)

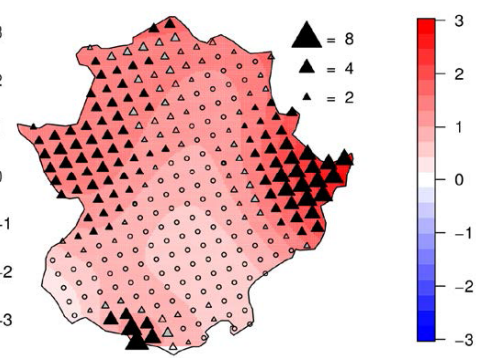
Trends in WE intensity (gridded data)

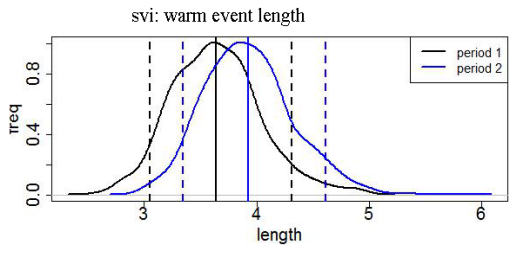
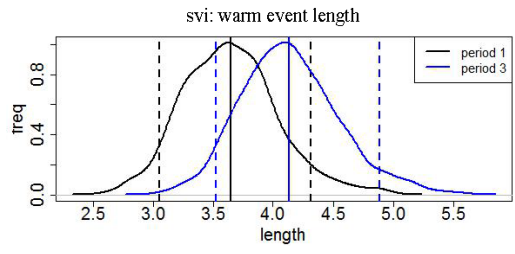
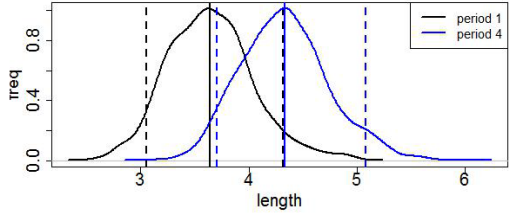
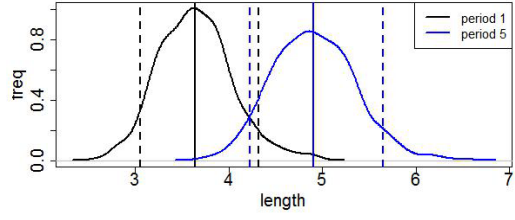

Trends in the 25th percentile-observational data 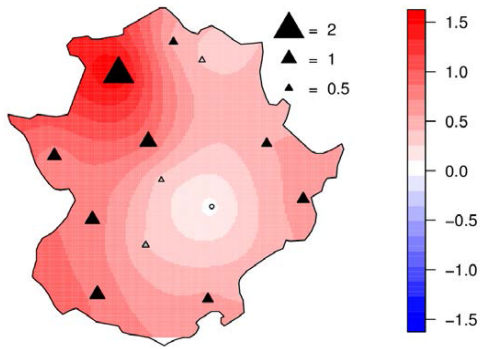 Trends in the 25th percentile-gridded data 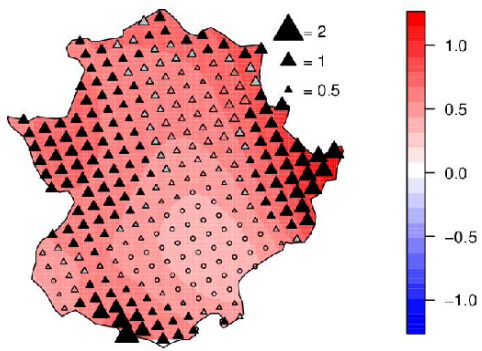

Trends in the 75th percentile-observational data 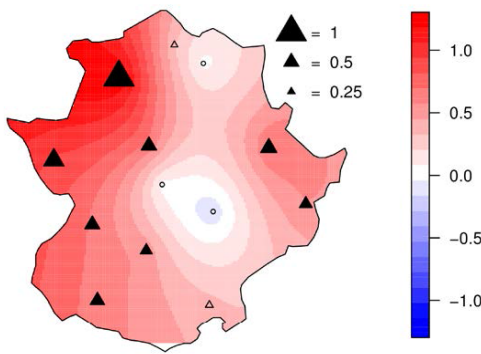 Trends in the 75th percentile-gridded data 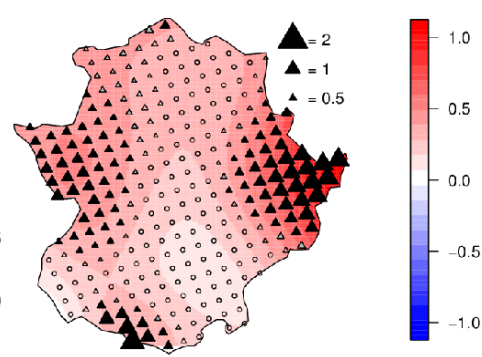

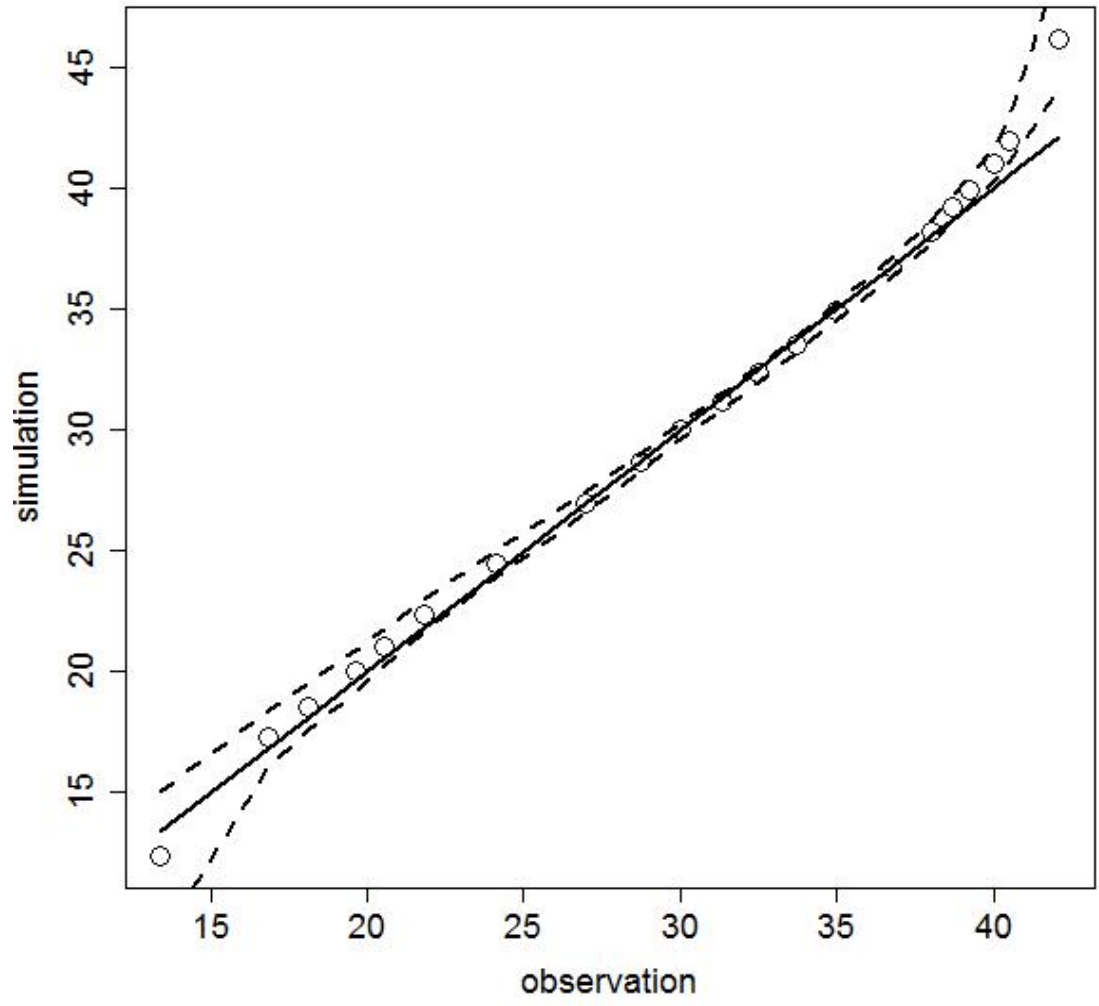

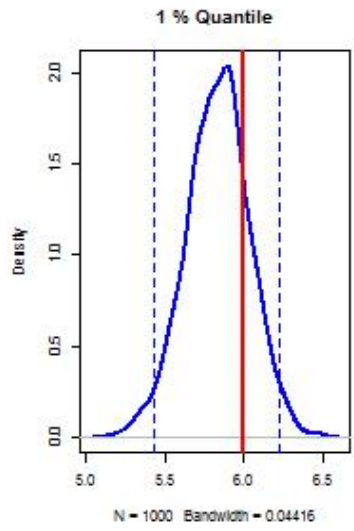 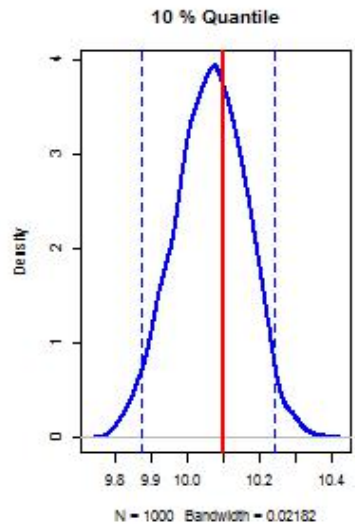 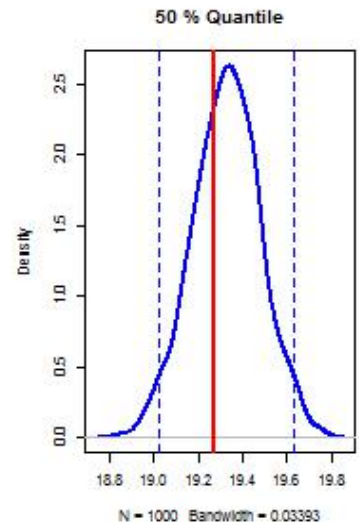 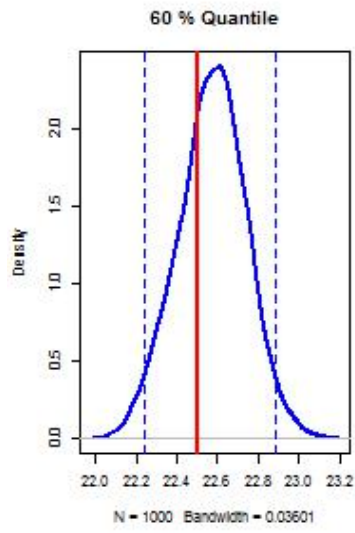 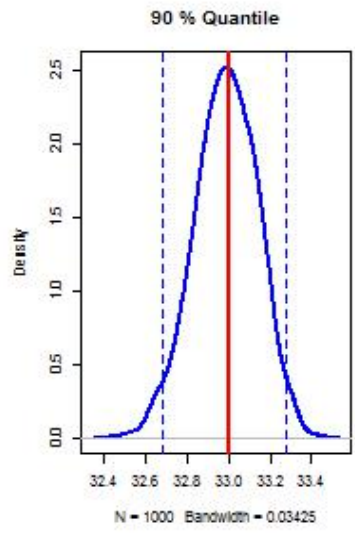 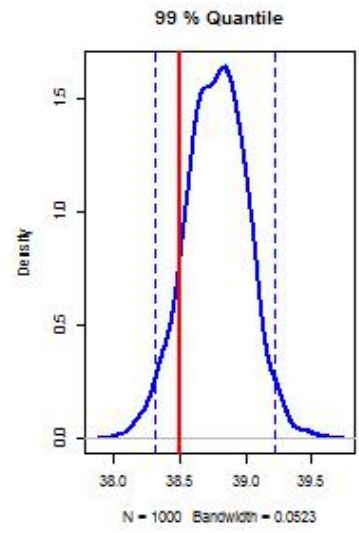

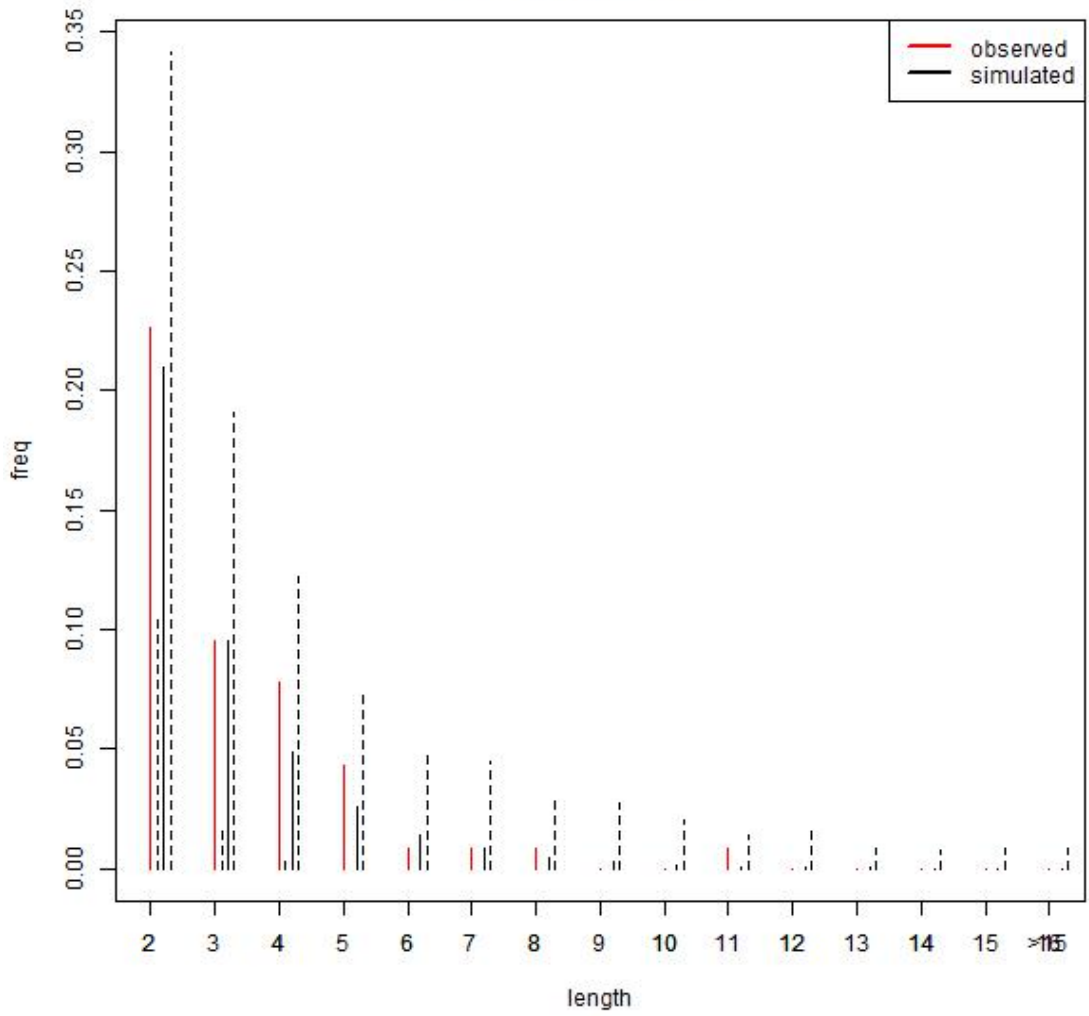